\documentclass[fleqn,usenatbib]{mnras}

\usepackage{newtxtext,newtxmath}

\usepackage[T1]{fontenc}

\DeclareRobustCommand{\VAN}[3]{#2}
\let\VANthebibliography\thebibliography
\def\thebibliography{\DeclareRobustCommand{\VAN}[3]{##3}\VANthebibliography}

\usepackage{graphicx}	
\usepackage{amsmath}	
\usepackage{graphicx}
\usepackage{dcolumn}
\usepackage{bm}
\usepackage{amsmath}
\usepackage{hyperref}
\usepackage[utf8]{inputenc}
\usepackage{xcolor}
\usepackage{comment}
\usepackage{soul}

\def\d{{\rm d}}

\title[INTEGRAL constraints on light DM]{Constraints on light decaying dark matter candidates from 16 years of INTEGRAL/SPI observations}

\author[F. Calore et al.]{
F.~Calore, $^{1}$\thanks{calore@lapth.cnrs.fr} 
A.~Dekker, $^{2}$\thanks{a.h.dekker@uva.nl}
P.~D.~Serpico,$^{1}$\thanks{serpico@lapth.cnrs.fr}
T.~Siegert, $^{3}$\thanks{thomas.siegert@uni-wuerzburg.de}
\\
$^{1}$LAPTh,  CNRS, USMB,  F-74940 Annecy, France\\
$^{2}$GRAPPA Institute, University of Amsterdam, 1098 XH Amsterdam, The Netherlands\\
$^{3}$Institut f\"ur Theoretische Physik und Astrophysik, Universit\"at W\"urzburg, Campus Hubland Nord, Emil-Fischer-Str. 31, 97074 W\"urzburg, Germany
}

\date{Accepted XXX. Received YYY; in original form ZZZ}

\begin{document}
\label{firstpage}
\maketitle

\begin{abstract}
We apply the recently developed analysis of 16 years of INTEGRAL/SPI data including a dark matter spatial template to derive bounds on  dark matter candidates lighter than WIMPs (like sterile neutrinos or axion-like particles) decaying into line or continuum electromagnetic final state channels. The bounds obtained are the strongest to date for dark matter masses between $\sim $60 keV and $\sim$16 MeV experiencing two-body decays producing photon lines.
\end{abstract}

\begin{keywords}
Astroparticle physics -- gamma rays -- dark matter 
\end{keywords}


\section{Introduction}
While gravitational evidence for the dark matter (DM) phenomenon has been steadily accumulating via astrophysical and, above all, cosmological observations, its nature remains an open problem to date. Most efforts have been devoted to identify {\it weakly interacting massive particles} (WIMPs) in the GeV-TeV mass range via observations of their putative annihilation into standard model (SM) byproducts ({\it indirect detection}), recoils in shielded underground detectors ({\it direct detection}), or missing energy in colliders, with negative results~\citep{Bertone:2004pz}.

There is no deep or fundamental reason, however, why the DM---if it is a particle at all---should lie in the electroweak mass range and should have electroweak-strength couplings with the SM. Alternatives to the WIMP paradigm abound, although they have been less thoroughly explored both theoretically and phenomenologically. Often, moving outside the WIMP class requires looking at {\it non-thermal} relics, with less generic signatures and diminished chances of discovery via  recoils or collider probes due to lower couplings (the so-called {\it Feebly Interacting Massive Particle}, or FIMPs~\citep{Bernal:2017kxu,Agrawal:2021dbo} belong to this class). 
Among the broad class of new feebly interacting particles, two very relevant DM candidates are 
axion-like particles (hereafter ALPs) and sterile neutrinos. 
ALPs are light pseudo Goldstone bosons unrelated to
the strong CP problem and arise in several beyond-the-Standard-Model particle physics theories, see e.g.~\cite{Jaeckel:2010ni}. 
Sterile neutrinos are fermion singlets of the Standard Model gauge group, coupled to the visible sector via the mass-mixing mechanism, whose nature may be directly linked to the generation of neutrino masses \citep{Drewes:2013gca}.

For masses around the MeV scale, indirect signals may show imprints in observationally challenging and relatively unexplored windows of the electromagnetic spectrum, the hard X-ray/soft $\gamma$-ray range~\citep{Essig:2013goa}. High statistics and large fields-of-views are particularly useful to improve the sensitivity to these candidates. 
 At MeV energies, the only instrument available today to tackle such searches is the coded-mask spectrometer telescope, SPI, aboard the INTEGRAL satellite~\citep{Winkler:2003nn,2003A&A...411L..63V}.
 Recently, in~\cite{Berteaud:2022tws} some of the authors of this work analyzed 16 years of data taken by SPI over the energy range 30 keV–8 MeV, extending the new measurement of the diffuse soft $\gamma$-ray emission with SPI to low energies between 0.5 and 8 MeV performed in~\cite{Siegert:2022jii}. A major innovation of~\cite{Berteaud:2022tws}  has been to perform a dedicated analysis of the diffuse MeV emission, including a DM-like spatial distribution as an independent template in the fit of SPI data, treating this component on the same footing as the known astrophysical components such as the Positronium (Ps) emission with a strong $\gamma$-ray line at 511 keV, and the diffuse inverse Compton (IC) scattering along the Galactic plane. No statistically significant detection of a DM-like signal was obtained, and a spectral analysis was performed to set bounds on DM candidates in the form of Hawking-evaporating, light, primordial black holes (PBHs)~\citep{Green:2020jor}. The same data can be however used to set constraints on other types of DM candidates, characterised by different emission mechanisms, notably decay, and other spectra possibly very different from the quasi-thermal one associated to PBHs. Here we extend the analysis to different spectral shapes associated to other classes of DM candidates. Doing that will also allow us to comment on aspects of the analysis outcome that depend on the spectral shape. We can anticipate that, just like for PBHs, for which the constraints obtained in~\cite{Berteaud:2022tws} are the strongest above $10^{17}$ g in mass, for decaying DM candidates we also obtain the most stringent bounds in a mass range from $\sim$60 keV to $\sim$16 MeV.
 
 This article is structured as follows: Sec.~\ref{methods} briefly describes our methodology, which strongly builds on the one adopted in~\cite{Berteaud:2022tws}. In Sec.~\ref{sec:general_limits}, we present our results and their impact on specific models, namely ALPs and sterile neutrinos. Sec.~\ref{sec:conclusions} is devoted to a final discussion and conclusions.

\section{Methodology}\label{methods}
Our analysis for both the putative DM signal and the astrophysical background closely follows~\cite{Berteaud:2022tws}, which we address the reader to for technical aspects. It suffices to remind here that, by fitting spatial templates to the SPI data in each energy bin, one can find the flux contribution (or upper limit) corresponding to each template, which in turn corresponds to different types of astrophysical or exotic contributions. 
No significant DM-like excess was detected in~\cite{Berteaud:2022tws}. The DM signal template is the integral along the line of sight of a typical Navarro-Frenk-White (NFW)~\footnote{Since the region of interest  in Galactic coordinates is rather broad, INTEGRAL-SPI angular resolution is not very sharp, and the signal from decaying DM is not as spiky as in the case of annihilating DM, the result is not affected by the extrapolation of the NFW profile towards the innermost region of the Galaxy, where it is also the most uncertain.} spatial distribution and, besides PBH evaporation, also represents the morphology expected for decaying DM. 
Hence, as done for the PBH evaporation rate in~\cite{Berteaud:2022tws}, we can set upper limits at 95\% confidence level on the DM decay rate. 
We perform  a maximum likelihood analysis using the 3ML package~\citep{Vianello:2015wwa}
to sample over the priors of the spectral parameters. 
In the fit, all energy bins are simultaneously fitted and we include the energy redistribution of SPI, 
by convolving the predicted model fluxes with the energy redistribution matrix.

As for the sampling over the DM parameter space (i.e.~mass and decay rate), we present a two-fold bound-setting 
procedure.
In what we call ``1D'' in the following, we set the upper limits separately for each mass value: For a fixed mass point in the DM parameter space, we obtain the upper limits on the decay rate by sampling it from a log-normal prior distribution. 
We set the 95\% upper limit corresponding to that mass value through the maximum likelihood analysis;
we then proceed to the next mass value, and repeat the procedure. 

This is also the customary procedure followed in the literature. For comparison purposes, we also set bounds in a ``2D''  procedure, where the sampling is done on the two independent parameters (mass and decay width) simultaneously.

For each DM sampling mode (1D vs 2D), we perform two different spectral analyses which differ 
in the choice of the spectral data-set and of the spectral model components (see Sec.~\ref{sec:conclusions} for a more detailed discussion). 
i) In a more global spectral analysis, we fit the total SPI inner Galaxy diffuse spectrum (see Fig.~2 left in~\cite{Berteaud:2022tws}) to a spectral model 
where the DM component is fitted simultaneously with other astrophysical components, as done in~\cite{Berteaud:2022tws} for the PBH case. For the latter, we include: 1)  A  cutoff power-law component corresponding to the population of unresolved point sources, mostly cataclysmic variables; 2) a line at 511 keV
with free normalisation to account for  positron annihilation with a free fraction of ortho-Ps for the continuum emission below 511 keV; 3) the decay line from $^{26}\mathrm{Al}$ at 1809 keV, with a free normalization\footnote{ Additional but much weaker lines may contribute to the Galactic spectrum, such as at 1173 and 1332 keV in the case of $^{60}\mathrm{Fe}$ ($<20\,\%$ of the 1809 keV flux), and at 1275 keV from $^{22}\mathrm{Na}$ (broadened, $<10\,\%$ of the 1809 keV flux). Our limits around these energies are therefore more conservative as they do not include these lines. Instrumental background lines are taken into account completely by the background modelling method \cite{Siegert2019_SPIBG}.}; 4) a power-law component with free amplitude and index for the dominant inverse Compton emission spectrum.
ii) A simplified analysis, where we consider, as spectral data points, the upper limits on the flux attributed to a DM-like template in the SPI template analysis (see Fig.~2 right in~\cite{Berteaud:2022tws}).
In this case, we compare the predicted DM flux with the flux upper bounds obtained on the DM-like template.
Whenever only residuals of a ``conventional'' analysis are available, which is especially true at MeV energies where the raw data analysis is challenging, the latter procedure is the typical one followed in the literature on DM bounds.

For the DM signal, we ignore the extra-galactic contribution since the SPI analysis is insensitive to isotropic contributions (in this respect, see~\cite{Iguaz:2021irx}).  The flux from DM decays thus reduces to the Galactic contribution given as
\begin{equation}
    \frac{\d\Phi}{\d E} = \frac{\Gamma}{4 \pi m_{\rm{DM}}} \frac{\d N_{\rm{decay}}}{\d E} \,  D,
    \label{eq:flux}
\end{equation}
 where $\Gamma$ is the DM decay rate, $m_{\rm{DM}}$ the DM mass and $\d N_{\rm{decay}}/\d E$ the energy spectrum per decay of the DM particle. Although  in the following we will limit ourselves to a single dominant decay channel, the generalisation to multiple channels  simultaneously present is straightforward: One should simply replace $\Gamma\,\d N_{\rm{decay}}/\d E \to \sum \Gamma_i\,\d N_{i}/\d E$ in eq.~(\ref{eq:flux}), with $\Gamma_i$ the width of the $i-$th channel (such that $\Gamma=\sum \Gamma_i$) and $\d N_{i}/\d E$ the associated spectrum.
The $D$-factor in eq.~(\ref{eq:flux}) describes the DM density profile $\rho$ of the Milky Way halo integrated over the line of sight $s$,
\begin{equation}
    D = \int \d s \, \rho(r(s,l,b))\,.
\end{equation}
We consider the spherically symmetric NFW density profile~\citep{Navarro_1997} with scale radius of $r_s=9.98$~kpc and corresponding density $\rho_s=2.2\times 10^{-24}$~g/cm$^3$~\citep{Karukes:2019jwa}. 
The $D$-factor is obtained around the Galactic halo within $-47.5^{\circ}\leq l\leq47.5^{\circ}$ and $-47.5^{\circ}\leq b\leq47.5^{\circ}$, corresponding to the region of interest of our spatial template, and we find $D=1.289\times 10^{29}$~keV/cm$^2$. 

The DM model dependence enters via the expected spectrum. We discuss here two types of decay channels,
\begin{equation}
\begin{split}
    \rm DM &\rightarrow \gamma + \gamma \\ 
    \rm DM &\rightarrow e^+ + e^- + \gamma \\
\end{split}
\end{equation}
whose width are denoted $\Gamma_{2\gamma}$ and $\Gamma_{\rm FSR}$ (for {\it Final State Radiation}), respectively. They are representative spectra for a large class of candidates~\citep{Essig:2013goa}.

The width $\Gamma_{2\gamma}$ is associated to the monochromatic spectrum, leading to a line-like signal at half the DM mass
\begin{equation}
    \frac{\d N_{\rm{decay}}}{\d E} = 2\delta \left(E-\frac{m_{\rm DM}}{2} \right).
\label{eq:axEnergySpectrum}
\end{equation}
To the above spectrum, we apply a Gaussian smoothing to take into account the energy resolution of SPI~\cite{Siegert_2022}.
We consider for this channel a DM mass range of 60 keV -- 16 MeV.

The width $\Gamma_{\rm FSR}$  is instead associated to the spectrum~\citep{Siegert_jan_2022}
\begin{equation}
    \frac{{\rm d}N_{\rm{decay}}}{{\rm d}E} \simeq \frac{\alpha}{2\pi}\left[ \frac{m_{\rm DM}^2 + (m_{\rm DM}-2E)^2}{m_{\rm DM}^2 E} \ln \left(\frac{m_{\rm DM}(m_{\rm DM}-2E)}{m_e^2} \right) \right],\label{dNdEFSR}
\end{equation}
with $m_e$ the electron mass and $\alpha$ the fine structure constant. The FSR channel opens at $m_{\rm DM} \gtrsim$ 1~MeV, and we therefore consider a DM mass range of $1.1$--$200$~MeV. 
In the following section, besides the phenomenological results on $\Gamma_{2\gamma}$ and $\Gamma_{\rm FSR}$  vs. $m_{\rm DM}$, we will also comment on their relevance for specific particle physics models for DM. 

\section{Results and implications}\label{sec:general_limits}
\begin{figure*}
    \centering
    {{\includegraphics[width=0.45\textwidth]{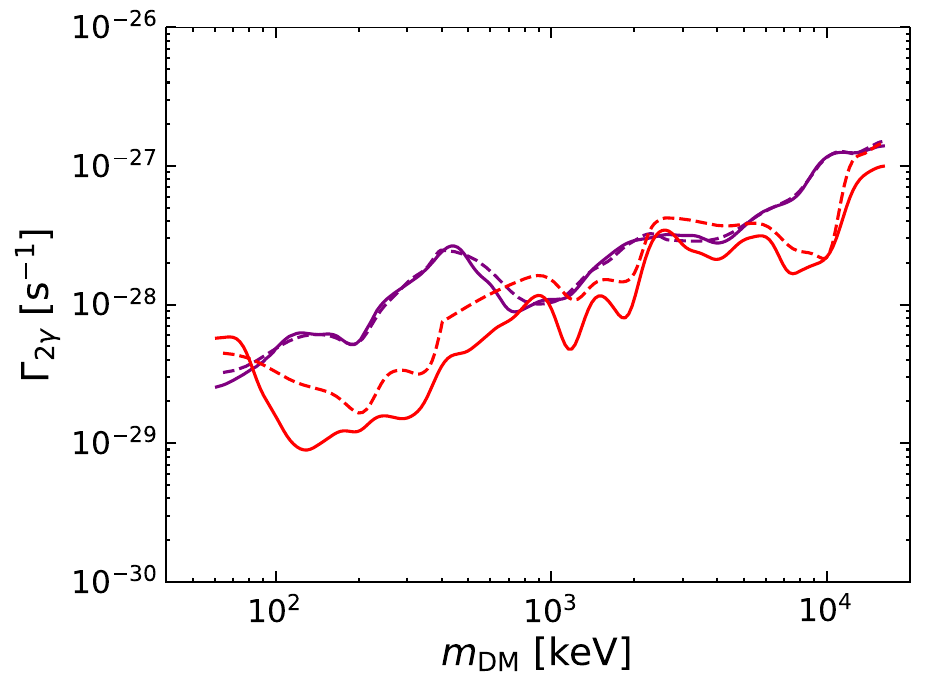} }}
 \qquad
    {{\includegraphics[width=0.45\textwidth]{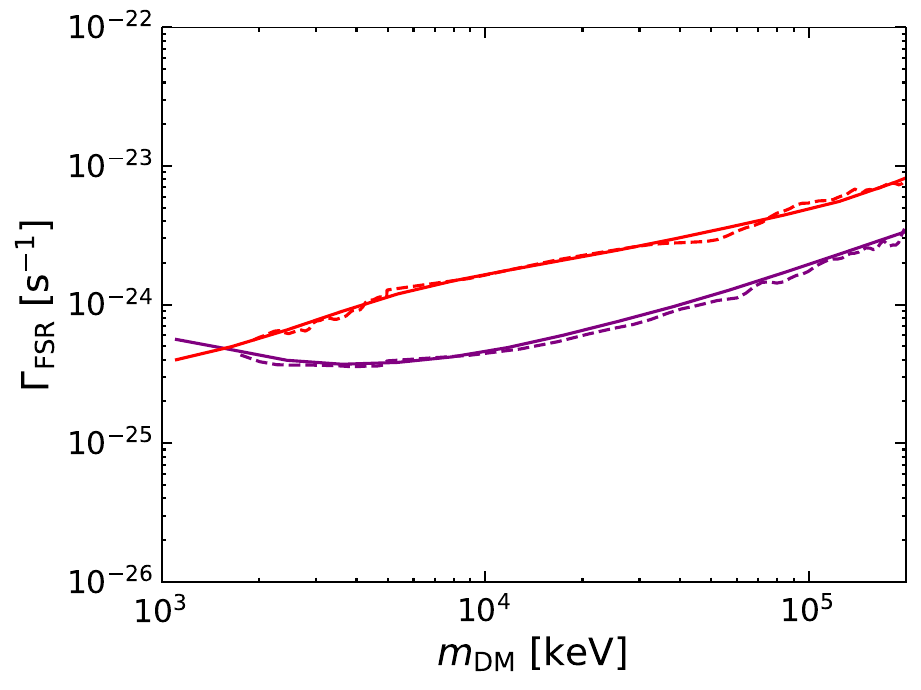} }}
  \caption{Upper limits on the DM decay rate at 95\% C.L. for the $2\gamma$ (left) and the FSR channel (right). Regions above the lines are excluded. In red (purple) we report the result of the total (NFW residuals only) spectral analysis by considering the 1D (2D) parameter space, shown as the solid (dashed) line.
  }\label{fig1}
\end{figure*}
In Fig.~\ref{fig1} we show our results for the $2\gamma$ (left) and the FSR channel (right), respectively. In solid (dashed) we report the result for a 1D (2D) analysis, in red (purple) we report the result for a total (NFW residuals only) spectral analysis. Values of the widths {\it above} the curves are excluded at 95\% C.L.

We note how the mass-range constrained by the FSR analysis is significantly higher than the one constrained by the line decay, since photons in the FSR case have a typical energy significantly lower than the mass of the DM, while the energy of the line signal is half the DM mass.

Additionally, for the FSR case the dependence of the bound from the mass is much smoother, since each mass contributes a broader spectrum
than the line one. Indeed, the line bounds are rather ``noisy'', since the signal only contributes to one energy bin (modulo instrumental effects implying energy redistribution), and stochastic fluctuations of the residuals from bin to bin affect the bounds discontinuously.

Also, the overall strength of the bounds is much higher for the monochromatic case. This is due to two major reasons: i) The much higher branching ratio in photons for the monochromatic case, since the FSR suffers from the suppression factor $\alpha/(2\pi)\simeq{\cal O} (10^{-3})$ in Eq.~\ref{dNdEFSR}. ii) The fact that the radiated spectrum is further distributed over a broad range (and thus more easily hidden within the error bars of the residuals) in the FSR case. 

Overall, the 2D analysis leads to bounds comparable to the 1D case, suggesting that there is little degeneracy between mass (affecting both normalisation and spectrum) and lifetime (only controlling the normalisation of the signal). Some modest differences are easier to see for the monochromatic case, of course, since only a single bin contributes to the bound. A technical caveat is that the 2D analysis is more prone to spurious features induced by under-sampling of the two dimensional parameter space. To obtain reliable bounds over several decades of mass requires particular care with the sampling algorithm, and it is usually more numerically expensive. To conform with the standard practice in the literature, in the following we will adopt the 1D bounds as our benchmarks.

On the other hand, performing an analysis of all spectral components or an analysis limited to the NFW residuals leads to bounds (red vs purple lines, respectively) that can differ by a factor up to $\sim 9.4$ and $\sim 3.6$ for the $2\gamma$ and FSR channel, respectively. Which type of analysis leads to stronger constraints is not obvious to predict a priori, since the result depends on the correlations (which can be positive or negative) between the different spectral components. This is particularly tricky to get an intuitive understanding of in the case of multiple components, since visualizing correlations in a D-dimensional space with $D>3$ is rather complicated. What is easier to understand is that the astrophysical component mostly degenerate with the DM one crucially depends on the spectrum of the DM signal.

\begin{figure*}
    \centering
    {{\includegraphics[width=0.45\textwidth]{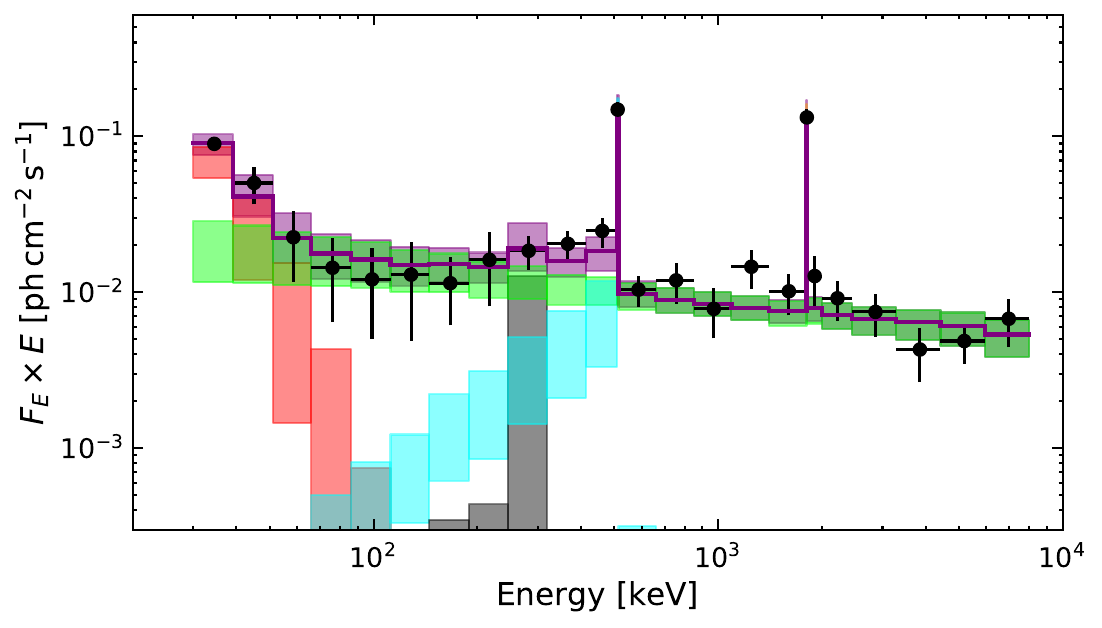} }}
 \qquad
    {{\includegraphics[width=0.45\textwidth]{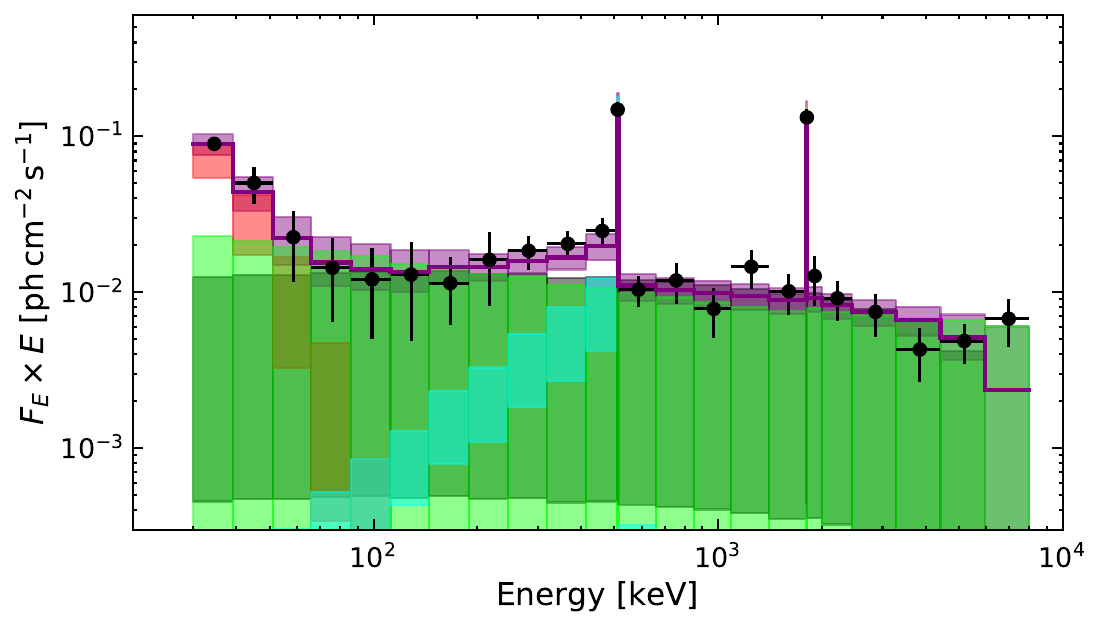} }}
  \caption{Spectral data points (black) and the combination of fluxes leading to the $95\%$ upper limit: Unresolved astrophysical point sources are indicated in red histograms, the positronium emission in cyan histograms, the diffuse inverse Compton scattering in green histograms, the DM contribution in grey histograms. The sum of all
  best-fit components is displayed in purple.
The following DM models are adopted: $2\gamma$ channel with $m_{\rm DM}=500$~keV (left panel), and FSR channel with $m_{\rm DM}=12$~MeV (right panel).
}\label{fig2}
\end{figure*}

In Fig.~\ref{fig2} we show the combination of fluxes leading to the $95\%$ confidence level bounds, as well as the spectral points, reported in black: Following the same color-code of~\cite{Berteaud:2022tws},  the unresolved point sources are indicated in red, the Ps emission in cyan, the diffuse IC scattering in green (in addition, there are nuclear lines not visible in the plot, to avoid clutter). The barely excluded DM signal is represented in grey. We illustrate the $2\gamma$ channel with $m_{\rm DM}=500$~keV in the left  panel and the FSR channel with $m_{\rm DM}=12$~MeV  in the right panel.  In the former case, we see that the line spectrum is quite distinct from the other components; if the astrophysical model falls somewhat short of the counts in that bin, introducing the DM component typically improves the situation in a comparable way for both the global and NFW-only analysis and the limits on DM degrade (this is what happens around a mass of 3 MeV in Fig.~\ref{fig1} left panel, for instance, to be compared with the undershooting of the model at energies around 1.5 MeV in Fig.~\ref{fig2}). If the astrophysical model already overshoots somewhat the data, then the inclusion of the DM must be accompanied by a downscaling of the dominating astrophysical background there, which leads to tighter bounds on DM because the downscaling strengthens the tension due to other bins. This is what happens, for instance, for a mass of 8-10 MeV in the left panel of Fig.~\ref{fig1}, i.e. energies around 4-5 MeV in the left panel of Fig.~\ref{fig2}, between the DM and the inverse Compton component. This correlation makes the effective room available for a DM residual smaller (hence the global bound tighter) than the naive NFW analysis in that bin would suggest. 

Conversely, let us focus on what happens with the FSR channel in the right panel shown in Fig.~\ref{fig2}, where $m_{\rm DM}=12$~MeV: The DM spectrum is extremely similar to the inverse Compton one. Indeed, the DM spectrum can be approximated as a power-law up to the cut-off with spectral index -1.1, while the inverse Compton spectrum is best fitted with a power-law index of -1.2. One can add a DM signal while simultaneously downscaling the inverse Compton contribution with little consequence for the global fit. Hence, in a global fit approach the actual room available for DM is bigger than the NFW-only residual analysis would suggest, and the bounds are tighter for a NFW-only analysis than for a global approach. These trends just described closely match what seen in Fig.~\ref{fig1}, right panel, in the high-mass range. 

On the other hand, if going to lower masses, for the line case one eventually populates the lowest/two lowest energy bins, where the DM spectrum becomes less distinguishable from the unresolved source contribution: The same phenomenon just described for the $m_{\rm DM}=12$ MeV FSR spectrum is now operational, and the NFW analysis yields stronger bounds than the global analysis. The opposite behaviour manifests for the FSR case at the lowest masses: In this case (not shown) multiple components are now relevant, the degeneracy between DM and inverse Compton spectra is less and less true, and the outcome is to make the global fit constraint slightly stronger.

Qualitatively, for rather distinctive DM spectra (such as the blackbody-like characteristic of PBHs considered in~\cite{Berteaud:2022tws}) the simplified comparison with the DM-like residuals is expected to lead to conservative bounds rather than more realistic ones, while constraints on DM spectra almost degenerate with some of the astrophysical components are expected to be artificially stronger in a simplified analysis. Whenever possible, our results suggest to resort to global fits in order to have a realistic assessment of the actual bounds on DM. 
This is also the benchmark assumed in all our following results.

\subsection{Impact on specific models}

\subsubsection{Axion-like particles} \label{sec:alps}

\begin{figure}
    \centering
    \hskip7.mm
    \includegraphics[width=0.5\textwidth]{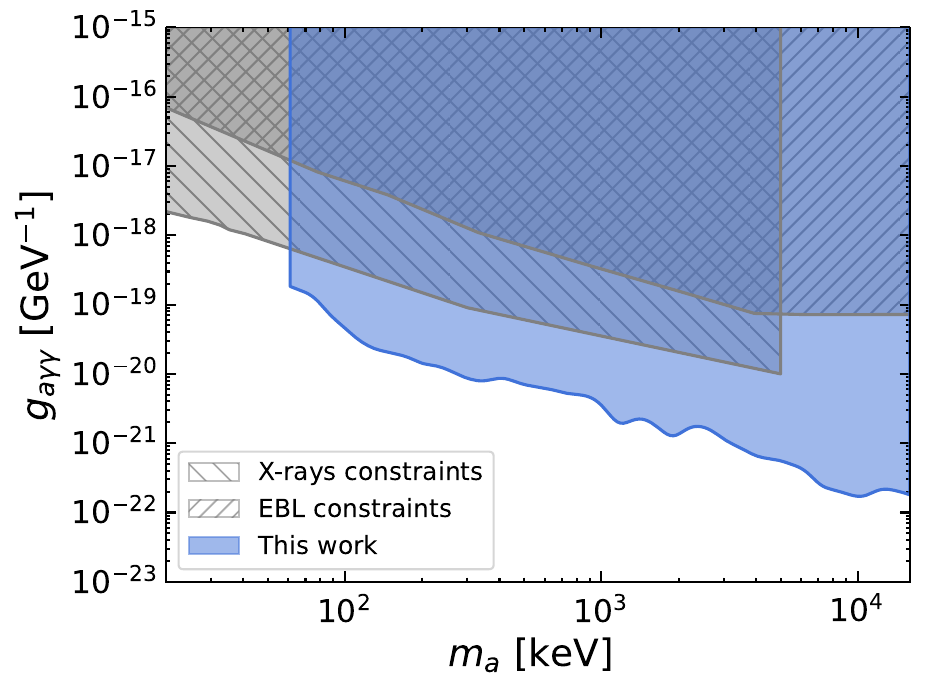}
\caption{Upper limit on the ALP-to-photon coupling at 95\% CL as a function of the ALP mass. Grey hatched regions represent previous X-ray and extragalactic background light constraints from~\protect\cite{Cadamuro:2011fd,AxionLimits}.}   \label{fig3}
\end{figure}

ALPs, whose defining coupling to two photons is $g_{a\gamma\gamma}$, can provide DM candidates in the MeV mass range~\citep{Arias:2012az}. They  are metastable, decaying into two photons ($a\rightarrow \gamma \gamma$) with a lifetime also depending on the ALP mass $m_a$. The  bounds previously obtained for the $\Gamma_{2\gamma}$ apply directly to this case, with $\Gamma_{2\gamma}$ writing in terms of ALP parameters as~\citep{Higaki:2014zua}
\begin{equation}
     \Gamma_{2\gamma}=\frac{g_{a\gamma\gamma}^2m_a^3}{64\pi} =0.755\times 10^{-30}\left(\frac{g_{a\gamma\gamma}}{10^{-20} \, \rm{GeV}^{-1}}\right)^{2}\left(\frac{m_a}{100 \, \rm{keV}} \right)^3\, {\rm s}^{-1}\,.
\end{equation}
In the range of ALP masses covered by our analysis, we obtain the strongest constraints available in the literature, as shown in Fig.~\ref{fig3} as the blue shaded area. The hatched grey areas represent currently excluded regions based on the extragalactic background light and X-ray observations according to~\cite{Cadamuro:2011fd,AxionLimits}. 

\begin{figure}
    \centering
    \hskip7.mm
    \includegraphics[width=0.5\textwidth]{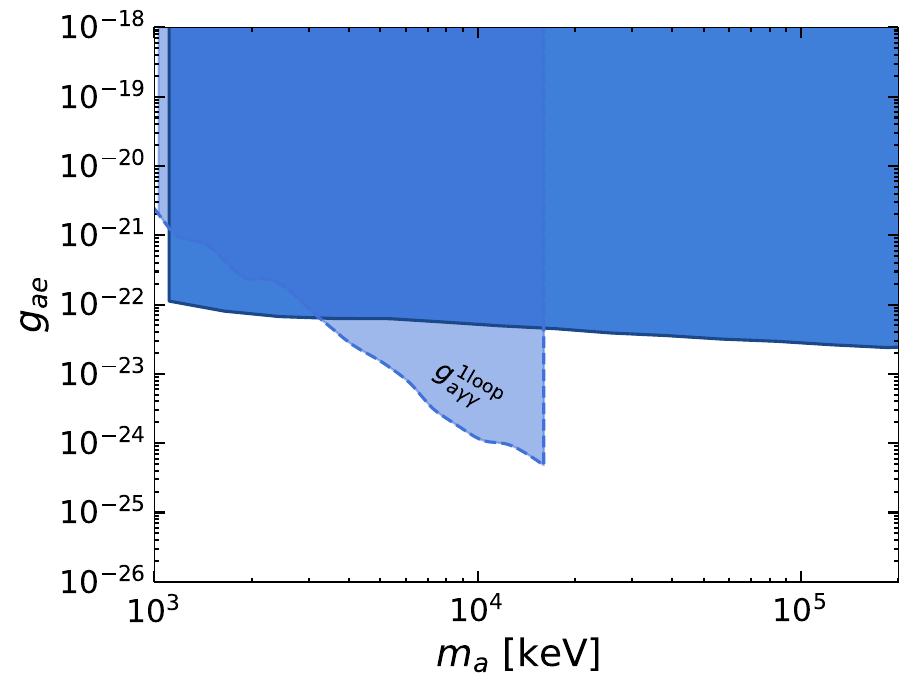}
\caption{Upper limit on the ALP-to-electron coupling at 95\% CL as a function of the ALP mass. Solid line: Direct bound from the FSR analysis. Dashed line: Expected indirect bounds from loop induced line contribution, valid at the order-of-magnitude level in absence of fine-tuned cancellations. }   \label{fig4}
\end{figure}

For $m_a >$ MeV, also the $e^\pm$ decay channel opens up, according to 
\begin{equation}
\Gamma\left(a \rightarrow e^{+} e^{-}\right)=\frac{g_{a e}^{2}}{8 \pi} m_{a}\left(1-\frac{4 m_{e}^{2}}{m_{a}^{2}}\right)^{1 / 2}\,,
\end{equation}
where $g_{ae}$ is the pseudoscalar coupling $-i g_{ae} a \bar{\psi}_e \gamma_{5} \psi_e$, and we naively expect $g_{ae}=m_e/f_a$ in QCD axion models, where $f_a$ is the Peccei-Quinn spontaneous symmetry breaking scale.  Although direct FSR bounds apply to this range of masses, it is worth noting that in presence of an ALP coupling with electrons, barring extremely fine-tuned cancellations with other contributions, we expect a loop induced
$g_{a\gamma\gamma}^{1\,{\rm loop}}\approx \alpha g_{ae}m_a^2/(12\pi m_e^3)$~\citep{Pospelov:2008jk,Ferreira:2022egk}, which typically implies much tighter bounds on $g_{ae}$ from the monochromatic signal previously discussed whenever available, as illustrated in Fig.~\ref{fig4}. 

\subsubsection{Sterile neutrinos} \label{sec:sterile_nu}

Another example of a DM model that leads to a monochromatic photon line in the keV-MeV mass range is a sterile neutrino~\citep{Drewes:2016upu}. Such a DM candidate has several decay modes, one of which at least is electromagnetic in nature and always kinematically allowed: 
It is the final state $\nu_a\gamma$, with $\nu_a$ one of the lighter, mostly active neutrino states. Since sterile neutrinos are much heavier than active ones~\footnote{Phenomenologically, the mixing angles describing the fraction of the heavy state $\nu_4$ in  active flavour eigenstates must be very small, so that the sterile flavour state $\nu_s$ almost coincides with the heavy mass state $\nu_4$. In the following, with a little abuse of notation we will thus denote $\nu_s$ as both the sterile flavour eigentstate and the  mass eigenstate $\nu_4$.}, the photon energy is $E_{\gamma}=m_{\nu_s}/2$. The corresponding energy spectrum per sterile neutrino decay only differs from the previously considered one by a factor 2:
\begin{equation}
    \frac{\d N_{\rm{decay}}}{\d E} = \delta \left(E-\frac{m_{\nu_s}}{2} \right).
\label{eq:EnergySpectrum}
\end{equation}

\begin{figure}
    \centering
    \hskip7.mm
    \includegraphics[width=0.5\textwidth]{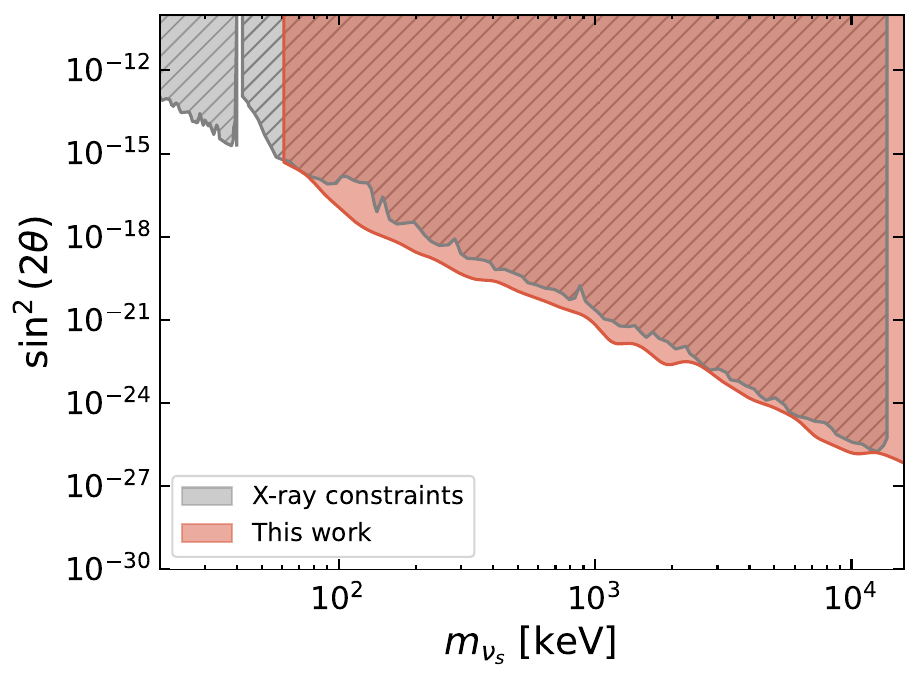}
\caption{
Upper limit on the mixing angle at 95\% CL as a function of the sterile neutrino mass. Grey hatched region represents previous X-ray constraints from~\protect\cite{Boyarsky:2007ge,Roach:2022lgo}.
}    \label{fig5}
\end{figure}

The width for this channel is given as (e.g.~\cite{Bezrukov:2009th})
\begin{equation}
    \Gamma_{\nu\gamma}\simeq \frac{9\alpha G_F^2 m_s^5\sin^2(2\theta)}{1024\pi^4} \simeq 1.36 \times 10^{-29} ~ {\rm s}^{-1} \left[ \frac{ \sin^2(2\theta)}{10^{-7}} \right]\left(\frac{m_{\nu_s}}{1~ {\rm keV}} \right)^5,\label{gammanuwidth}
\end{equation}
where $\theta$ is the mixing angle between the sterile and active neutrino.  We can thus recast the bounds in the $m_{\nu_s}-\sin^2(2\theta)$ plane by simply loosening the bounds on $\Gamma_{2\gamma}$ previously obtained by a factor 2, and equating to eq.~\ref{gammanuwidth}. Our bounds are reported in Fig.~\ref{fig5}, and compared with previous X-ray constraints from~\cite{Boyarsky:2007ge,Roach:2022lgo}, resulting tighter than those. 

A few comments are in order: We prefer not to recast the bounds on sterile neutrinos in terms of neutrino lifetime. This is because the neutrino lifetime is not given by $1/\Gamma_{\nu\gamma}$ and the conversion may be confusing. There is always at least a 3$\nu$ decay mode which is open, and it is dominating over the radiative mode by two orders of magnitude. 
At high masses, $m_{\nu_s}>2m_e\gtrsim 1$~MeV, sterile neutrinos can additionally decay to an electron-positron pair and active neutrino ($\nu_s \rightarrow \nu_a e^+ e^-$). The width of this mode depends on the mixing being with $\nu_e$  or $\nu_{\mu,\tau}$, but it is ${\cal O}$(10\%$\div$ 30\%) and always at least one order of magnitude larger than the $\nu\gamma$ mode. A more complete analysis of sterile neutrino bounds at $m_{\nu_s}>2m_e$ would require combining a FSR mode with the monochromatic mode channel, but based on our results of Fig.~\ref{fig1} we can still make the reasonable assumption that the monochromatic bound dominates.  In any case, our approximation means that the bound reported in Fig.~\ref{fig5} is, if anything, mildly conservative at high masses.

At a more technical level: We are implicitly (and conservatively) assuming that the sterile neutrino is a Dirac particle. For a Majorana state, all (exclusive) widths are the same, but also the lepton-number conjugated ones are allowed,  hence (inclusive) widths are double the ones reported above, for the same coupling. For instance, the channel $\nu_s\to \nu_a \gamma$ is accompanied by the channel 
$\nu_s\to \bar{\nu}_a \gamma$ having the same rate. 

\section{Discussion and conclusions}\label{sec:conclusions}
In this work, we have applied a recent analysis of 16 years of INTEGRAL/SPI data to derive bounds on decaying DM channels. 
This works builds on~\cite{Berteaud:2022tws} where, for the first time, a DM morphological template has been included as independent model component and simultaneously fit together with various astrophysical background components to 
properly assess the significance of a DM-like signal.
In the absence of a significant DM-like excess,~\cite{Berteaud:2022tws} used spectra emitted by PBH DM candidates via Hawking evaporation to derived the strongest bounds on the DM fraction of light PBHs in the mass range $10^{17} - 10^{18}$ g.

We have extended in this respect the work of~\cite{Berteaud:2022tws}, and applied the same procedure to different DM spectra corresponding to some popular models of non-WIMP DM in the keV-MeV range, such as sterile neutrinos and axion-like particles.
We have considered two main decay channels for production of MeV photons: The decay into gamma-ray lines and the electromagnetic final state radiation emitted when the decay into electron and positron pairs is allowed.
We have 
obtained the strongest bounds to date to these decay widths in the part of the model mass space kinematically probed by the INTEGRAL/SPI data, i.e. 60 keV -- 16 MeV. 

Besides these results of physical relevance, the use of these two different type of decay spectra, a spectral line and a much broader final state radiation spectrum, has allowed us to discuss the role that the DM spectrum plays in deriving bounds in a multi-component fit. In particular, we have discussed how a global analysis of the whole spectrum or a simplified analysis of the residuals relative to a DM-like component can lead to bounds differing by factors of a few, and up to $\sim 9$. In particular, the simplified analysis can overestimate constraints when the DM spectrum is actually too degenerate to the one of an astrophysical component entering the global analysis, or underestimate it when the DM spectrum is instead quite distinctive and the astrophysical model already explains the data well, or mildly overshoots them. 

Our work highlights the relevance of further exploring the hard X-ray/soft $\gamma-$ray band for tightening the exploration of non-WIMP DM models, and calls for dedicated missions in this energy range, such as the accepted COSI MeV $\gamma$-ray satellite mission, to be launched by NASA in 2026 \citep{Tomsick2019_COSI}. In order to sample reliably the DM mass range beyond 10\,MeV, future space missions, such as ASTROGAM \citep{deAngelis2018_ASTROGAM} or AMEGO \citep{Kierans2020_AMEGO} will be required.
It also issues a warning to the community of the risks of (over)simplified analyses, if one aims at obtaining realistic and better than ``order-of-magnitude'' bounds on DM properties.

\section*{Acknowledgements}
We warmly thank J.~Berteaud and J.~Iguaz for initial help with the DM 3ML analysis.
A.D.~would like to thank LAPTh and the University of Amsterdam for supporting her visits.
We acknowledge funding by the ``Agence Nationale de la Recherche”, grant n. ANR-19-CE31-0005-01 (PI: F. Calore). 

\section*{Data availability}
The spectral data points and response files from INTEGRAL/SPI are available in an online repository (\url{https://zenodo.org/record/7984451}). 

\bibliographystyle{mnras}
\bibliography{references}

\end{document}